\documentclass[aps,twocolumn,prd,nofootinbib,showpacs]{revtex4}
\usepackage[pdftex,bookmarks]{hyperref}

\usepackage{amsfonts}
\usepackage{amsmath,amssymb}
\usepackage[T1]{fontenc}
\usepackage{mathrsfs}
\usepackage{epsfig,epstopdf}

\usepackage[absolute]{textpos}

\usepackage{graphics,wrapfig,times}
\usepackage{graphicx}

\usepackage{color}
\definecolor{D}{rgb}{0.00,0.17,0.48}
\definecolor{M}{rgb}{0.00,0.02,0.83}
\definecolor{L}{rgb}{0.58,0.79,1.00}
\definecolor{R}{rgb}{1,0,0}
\definecolor{B}{rgb}{0.00,0.00,0.00}
\definecolor{P}{rgb}{0.00,0.30,0.60}
\definecolor{W}{rgb}{1,1,1}

\TPGrid[40mm,40mm]{15}{25}  

 \makeatletter
 \newcommand{\pback}[1]{{
   \let\@rrow=\leftarrowfill
   \mathchoice{\AIN@stemPullBack{#1}{\@rrow}}{\AIN@stemPullBack{#1}{\@rrow}}
     {\AIN@indxPullBack{#1}{\@rrow}}{\AIN@indxPullBack{#1}{\@rrow}}}
   \vphantom{#1}}

 \newcommand{\AIN@stemPullBack}[2]{
   \vtop{\mathsurround=0pt
   \ialign{##\crcr$\textstyle{#1}\strut$\crcr
     \noalign{\kern-0.4ex\nointerlineskip}{\tiny#2}\crcr}}}

 \newcommand{\AIN@indxPullBack}[2]{
   \vtop{\mathsurround=0pt
   \ialign{##\crcr\hfil$\scriptstyle{#1}$\hfil\crcr
     \noalign{\kern+0.4ex\nointerlineskip}{\tiny#2}\crcr}}}

\def\bar{\overline}
\def\be{\begin{equation}}
\def\ee{\end{equation}}
\def\bea{\begin{eqnarray}}
\def\eea{\end{eqnarray}}
\def\ba{\begin{array}}
\def\ea{\end{array}}

\def\={\hateq}
\def\puto#1{\rlap{\raise.5ex\hbox{\char'27}}{#1}}

\newcommand{\nn}{\nonumber}
\newcommand{\half}{\frac{1}{2}}
\def\eth{\text{\dh}}
\def\thorn{\text{\th}}
\def\a{\alpha}
\def\b{\beta}
\def\c{\gamma}
\def\d{\delta}

\def\IH{\triangle}

\def\ta{{\tilde{a}}}
 \def\tb{{\tilde{b}}}

\def\tn{{\tilde{0}}}
\def\one{{\tilde{1}}}
 \def\tt2{{\tilde{2}}}
  \def\tk{{\tilde{k}}}

\def\.{\cdot}
\def\D{{\cal D}}

\def\L{{\cal L}}

\def\M{{\mathscr M}}
\def\kl{\kappa_{(\ell)}}

\def\Re{{\rm Re}}
\def\Im{{\rm Im}}
\def\l{\ell}

\def\be{\begin{equation}}
\def\ee{\end{equation}}
\def\bea{\begin{eqnarray}}
\def\eea{\end{eqnarray}}
\def\ba{\begin{array}}
\def\ea{\end{array}}

\def\R{{\mbox{\rm$\mbox{I}\!\mbox{R}$}}}

\def\up{\stackrel}

\newcommand{\eqhat}{\mathrel{\widehat\mathalpha{=}}}
\def\={\eqhat}

\begin{document}
\title{Gravitational radiation of generic isolated horizons}

\author{Yu-Huei Wu}
\email{yhwu@astro.ncu.edu.tw, yuhueiwu@hotmail.com}
\affiliation{Institute of Astronomy, National Central University,\\
     Chungli, Taiwan 320, R. O. C.}

\author{Chih-Hung Wang}
\email{chwang@phy.ncu.edu.tw}
\affiliation{Department of Physics, National Central University,\\
     Chungli, Taiwan 320, R. O. C.}

\begin{abstract}

From the similarity between null infinity and horizons, we show
how to set up proper frames near generic isolated horizons. The
asymptotic expansion and reference spin frame are used to study gravitational radiation near generic isolated horizons and it turns out that the news function appears on non-expanding horizon. We also verify that the surface gravity is constant on (weakly) isolated horizon. The corresponding conserved quantities and relevant
asymptotic symmetry groups which allow gravitational radiation of
generic isolated horizons are obtained from asymptotic expansion.

\end{abstract}

 \maketitle

\section{Introduction}


The boundary of black hole is defined as  \textit{a region of no
escape}. However, the definition of event horizon cannot give a
realistic description of how a black hole grows since it is too
global. The event horizon can be located only after we know the
global structure of space-time. Hence, the purpose of
generalization of the event horizon to the 'quasi-local' horizons
is to \textit{let the observer detect the horizon}. The stationary
horizon excludes the situation of radiation outside the horizon.
Therefore, Ashtekar et al \cite{Ashtekar99b} propose the idea of
the generic isolated horizon. It is less restrictive than the
usual Killing horizon. We do not need to require any symmetry or
assume that the space-time is globally stationary. If gravitational
collapse occurs, the final stage of the the black hole is isolated
and in the equilibrium state therefore will not radiate any more.
However, there might be some gravitational and matter fields radiation which is far away from the black hole.
The horizon will finally reach an equilibrium state and settle down
to an \textit{isolated horizon} (IH). The purpose of this framework is
to probe the properties of black hole which are themselves in
equilibrium but allows non-trivial dynamics in the exterior
region. It allows one to assign mass and angular momentum to the
black hole in terms of values of the fields on the horizon itself
without referring to null or space-like infinity. It also leads to
a generalization of the zeroth and first laws of black hole
mechanics.

It is expected that black holes are rarely in equilibrium in Nature. By using
generic isolated horizons as a basis, the ideas can be generalized to a
\textit{dynamical horizon} definition by a space-like hypersurface
rather than null hypersurface in the non-expanding horizon
definition. The horizon geometry of dynamical horizon is time
dependent and it allows a quantitative relation between the growth
of the horizon area and the flux of energy and angular momentum
across it \cite{Ashtekar02}.

We start from Ashtekar's most general definition of isolated
horizon called the \textit{non-expanding horizon} (NEH).
It requires the degenerate metric to be independent of time. If we
further require the extrinsic curvature (the rotation one form) to
be time independent then it gives the definition of
\textit{weakly isolated horizon} (WIH). Here, the NEH resembles
Killing horizon up to the first order and WIH further up to the
second order. In WIH, the black hole zeroth law holds. One cam further require the full derivative operator to be time
independent, and it yields the definition of IH.

Unlike Ashtekar's three dimensional analysis our work is a fully
four dimensional approach. This approach allows one to consider
the next order contribution from the neighborhood of the isolated
horizon. It allows us to calculate the amount of mass-energy flux
cross or near the horizon. Although there is no well-defined
mass or energy density (including gravitational field) in general relativity (GR),
it does have well-defined mass or energy associated to a two surface, i.e., quasi-locally. Unlike
Newtonian theory, it does not have an unique expression of quasi-local mass
or energy in GR \cite{WuWang-2008-a}.  We use a quasi-local
formula based on spinor fields to define the mass of a black
hole. Asymptotic expansions gives a way to study the geometry near
black holes or null infinity. Using the similarity to the asymptotic expansion
for the null infinity we can set up a frame, certain gauge choices
near the boundary of the horizon. Therefore, we can find the
asymptotic expansion of the Newman-Penrose (NP) coefficients with
respect to radius and compare this with the exact solutions we
know. The asymptotic expansion for the null infinity and the
horizon are quite different geometrically. As we approach null
infinity we consider an asymptotically flat space-time, however,
the approach near the horizon is \textit{not necessarily
asymptotically flat}. For the null infinity, we take the incoming
tetrad $n$ as a generator of null infinity that generates
different cuts with respect to different times. The outgoing
tetrad $\l_a$ can be chosen as tangent to null ray that can be
parameterized by using affine parameter $r$. On the other hand,
$\l_a$ can be the gradient of the surface of a constant retarded
time $u$. On the non-dynamical horizon, we take the outgoing
tetrad $\l$ as the generator of the horizon that can generate
different cross sections with respect to different advanced times,
say $v$. The ingoing tetrad $n_a$ can be chosen as tangent to a
null ray that can be parameterized by using the affine parameter
$r$. This light ray goes into the horizon. The tetrad $n_a$ can be
the gradient of the surface of a constant retarded time $v$. In
this paper, we use convention $(+---)$ and NP equations in p.
46-p. 50 of \cite{Chandrasekhar}.

\section{The generic isolated horizons \label{IH}}



Firstly, we consider 4-D space-time manifold $(\M, g)$ with 3-D sub-manifold
$(\Delta, q)$.

\paragraph{Definition.} $\Delta$ is called a \textit{non-expanding
horizon} (NEH) if
 (1) $\Delta$ is diffeomorphic to the product $S \times
\R$ where $S$ is a space-like two surface.
 (2) The expansion $\Theta_{(\l)}$of any null normal $\l$ to $\Delta$
vanishes, where the expansion is defined by $\Theta_{(\l)}=\half
q^{ab}\nabla_a \l_b$ with $q_{ab}$ the degenerate intrinsic metric
on $\Delta$.
 (3) Field equations hold on $\Delta$ and $T_{ab}$ is such that
$T^a_{b}\l^b$ is causal and future-directed on $\Delta$.

\smallskip

Let $\Delta$ denote the three surface which gives the NEH. The
pair $(q, \D)$ consist of the intrinsic metric $q$ and the induced
derivative $\D$ where $\D_a =\!\pback{\nabla_a} = $ on $\IH$. The
intrinsic metric $q_{ab}$ on $\IH$ has signature $(0,-,-)$. The
vectors $(m^a, \bar m^a, \l^a)$ span the tangent space to $\Delta$
with the dual co-frame given by the pull backs of $(n_a, m_a, \bar
m_a)$. The expansion of outgoing and incoming null rays is defined
by
$ \Theta_{(\l)} := \half q^{ab} \nabla_a \l_b = - \Re \rho,$ and
$\Theta_{(n)} := \half q^{ab} \nabla_a n_b =  \Re \mu,$
where $q^{ab}:= -m^a\bar m^b-\bar m^a m^b$ on the tangent space of
horizon. The twist on horizon is defined as
 $\omega^2_{twist}:= \half q^a\,_c\; q^b\,_d \; \nabla_{[a}\l_{b]} \nabla^{[c}\l^{d]}.$
The shear on horizon is defined as
$ |\sigma_{shear}| := [\half q^a\,_c\; q^b\,_d \nabla_{(a}\l_{b)}
\nabla^{(c}\l^{d)} - \Theta_{(\l)}^2 ]^\half.$  Since $\l$ is the
null normal of the null hypersurface, it implies the twist free. Moreover, the shear vanishes by using Raychaudhuri equation
\footnote{For the outgoing null geodesic $\l$, the
\textit{Raychaudhuri equation} can be written as
\bea \L_{\l} \Theta_{(\l)} = - \Theta_{(\l)}^2 -\sigma_{shear}\bar\sigma_{shear} +
\omega_{twist}^2 + {\kappa}_{(\l)} \Theta_{(\l)} -\Phi_{00}. \nn\eea} and the dominate energy condition. Therefore, the gauge conditions on NEH
are
\bea \kappa\hat{=}0, \sigma\hat{=}0, \rho\hat{=}0.
\label{shearfreeIH}\eea
From using these, there must exist \textit{a natural connection
one form} $\omega :=\omega_a d x^a$ on $\IH$ which can be obtained by
\be \label{omega}\D_a \l^b\hat{=}\omega_a \l^b. \ee
\footnote{The $\hat{=}$ represents equal on horizon and the arrow
$\!\pback{}$ refers to the pullback of the index to $\IH$.}
%
%
%
The \emph{surface gravity} $\kappa_{(\l)}$ is defined as
\be\label{kappa}\kappa_{(\l)}:=\omega_a \l^a 
\ee
on NEH $\Delta$ (measured by $\l$). Note that we do not have
an unique normalization for $\l$.  Under the scale transformation
$ \l \mapsto f \l$, we have
$ \omega\mapsto \omega+d \ln f $ and $ \kappa_{(\l)}\mapsto f
\kappa_{(\l)}+ f \L_\l \ln f $
which leaves Eq. (\ref{omega}) and Eq. (\ref{kappa}) invariant.


From (\ref{omega}), we get
\bea     \L_\l q_{ab} \hat{=} \!\pback{\L_\l g_{ab}} \hat{=}
q_{cb} \omega_a \l^c + q_{ac}\omega_b \l^c =0\eea
for any null normal $\l$ to $\IH$.  In fact, $\l$ is an asymptotic
Killing vector field as we approach the horizon even though
\textit{the space-time metric $g_{ab}$ may not admit a Killing
vector field in the neighborhood of $\IH$.}


The energy condition from the third point of definition then
further implies that $R_{ab} \l^b$ is proportional to $\l_a$
\cite{Ashtekar02}, that is
$ R_{ab} \l^a X^b\hat{=} 0, $
for any vector field $X$ tangent to $\IH$. We then have
\be  \Phi_{00} \hat{=}  \Phi_{01} \hat{=} \Phi_{10} \hat{=} 0. \ee
Because $\l$ is expansion and shear-free, it must lie along one of
the principal null directions of the Weyl tensor. From equation
(b) and (k) in P. 46 in \cite{Chandrasekhar},  we have:
\be \label{weyl}  \Psi_0 \hat{=} \Psi_1 \hat{=} 0.  \ee
The $\Psi_2$ is gauge invariant i.e., independent of the choice of
the null-tetrad $(n,m,\bar{m})$ on $\IH$. We have
 \bea d \; \omega \hat{=} 2 (\Im[\Psi_2])\; ^2\epsilon \label{domega}\eea
where $^2\epsilon$ is an area two form. The two form $d \omega$ can
also be written as
\bea  2\D_{[a} \omega_{b]} =  2 (\eth \pi -\bar\eth\bar\pi) m_{[a}
m_{b]}.\eea
$\Im[\Psi_2]$ plays the roles of gravitational contributions to
the angular-momentum at $\IH$. Ashtekar et al calls $\omega$ the
\textit{rotational 1-form potential} and $\Im [\Psi_2]$  the
\textit{rotational curvature scalar}.


Using Cartan identity $\L_v = d i_v + i_v d$ and
(\ref{domega}), the Lie derivative of $\omega$ with respect to
$\l$ is given by
\be\label{dkappa} \L_\ell \omega_a \hat{=} 2\Im(\Psi_2)\, \l^b\,
{}^2\!\epsilon_{ba} + \D_a (\l^b\omega_b)\ \hat{=} \D_a\kl. \ee
On NEH, the surface gravity  may not be constant. To obtain the zeroth law such that the surface gravity is constant, one may need a further condition, i.e., $ \L_\ell \omega_a=0$,  on NEH. It
motivates the definition of \textit{weakly isolated horizon}.


\smallskip
\paragraph{Definition.} A \textit{weakly isolated horizon} (WIH) is a NEH
with an equivalence class of null normals under constant
transformation. The flow of $\l$ preserves the rotation 1-form
$\omega$
$\L_\l \omega_a\hat{=}0 $,i.e., $[\L_\l ,\D] \l \hat{=}0.$
%


From (\ref{dkappa}), the condition of WIH basically preserves the
black hole zeroth law. Because $\l$ is tangent to $\IH$, the evolution equation is in
fact a constraint. See (B21) and (B22) of \cite{Ashtekar02}.
Therefore, given a NEH, we can select a canonical $[\l]$ by
requiring $(\IH, [\l])$ to be a WIH satisfying
\bea {\cal L}_{\l} \mu\hat{=}0\;\; \textrm{or }\;\; \dot\mu\hat{=}0.\eea
$\IH$ generically admits an unique $[\l]$ such that the incoming
expansion is time independent. This result will establish that a
generic NEH admits an unique $[\l]$ such that $(\IH, [\l])$ is a
WIH on which the incoming expansion $\mu$ is time independent.

\smallskip
\paragraph{Definition.} A weakly isolated
horizon $(\Delta, [\l ])$ is said to be \textit{isolated horizon}
(IH) if
$
 [{\cal L}_{\l}, \D] V \hat{=} 0,
$
for all vector fields $V$ tangential to $\Delta$ and all $\l \in
[\l ]$.
\medskip

From this definition, we have  $[\L_\l, \D] \l \hat{=} 0$ and
$[\L_\l, \D] n \hat{=} 0$. The first one gives the
surface gravity is constant by previous argument. So $\dot
\epsilon\hat{=} 0$. The second one gives $\dot \pi \hat{=} \dot
\mu \hat{=}\dot \lambda \hat{=} 0$.

\section{Asymptotic structure and coordinate transformations near generic isolated
horizons} 

\textbf{Frame setting, gauge choice and gauge conditions}

We choose the incoming null tetrad $n_a=\nabla_a v$ to be gradient of
the null hypersurface $v=const.$ and it gives $g^{ab} v_{,a} v_{,a}
=0$. We further choose $m, \bar m$ tangent to the two surface. These
gauge choices lead to
\be\ba{lll} \nu &=& \mu-\bar\mu = \rho-\bar\rho= \c +\bar\c = \pi-\a-\bar\b=0,\\
\pi &=& \bar\tau. \label{gauge-c}\ea \ee
From the definition of NEH, the gauge conditions are
\bea \kappa = \kappa_0 r' + O(r'^2),  \rho   =  \rho_0 r' + O(r'^2),  \sigma = \sigma_0 r' + O(r'^2),\nn\\
 \epsilon-\bar\epsilon=O(r'),\nn \eea
where $\rho_0: =  [\Psi^0_2- \bar\eth_0\bar\pi_0+\pi_0\bar\pi_0]$,  $\sigma_0: = [- \bar\eth_0\bar\pi_0+\pi_0\bar\pi_0]$. The rest of NP coefficients are $O(1)$.  
%
%
The Weyl tensor has the fall off (refer to equation (\ref{weyl}))
\bea \Psi_0 = O(r'), \Psi_1 = O(r')\eea
where $r'=r-r_\Delta$. In order to preserve orthogonal relation
$ \l^a n_a=1, m^a \bar m_a =-1, \l^a m_a= n^a m_a =0,$
we can choose the tetrad as
\bea \l^a=(1, U, X^3, X^4),\;\;
          n^a=(0,-1,0,0),\;\;
          m^a=(0,0, \xi^3,\xi^4). \nn\eea
%

We first expand NP spin
coefficients, tetrad components $U, X^k, \xi^k$ and Weyl
spinors $\Psi_k$ with respect to $r'$ and substitute them into NP equations to get following equations: 
%
%
%
 \be\ba{lll}
 \xi^k &=& \xi^{k0} + O(r'), \\
 U &=& 2 \epsilon_0 r' +  O(r'^2), \\
 X^k &=&  2(\pi_0\xi^{k0}+\bar\pi_0 \bar\xi^{k0})r'+ O(r'^2), \\
\Psi_0 &=& \half (- \eth_0 \Psi_1^0 + 4 \bar\pi_0 \Psi_1^0 - 3 \sigma_0 \Psi_2^0) r'^2+ O(r'^3)\\
\Psi_1 &=& (- \eth_0 \Psi_2^0 + 3 \bar\pi_0 \Psi_2^0) r'+ O(r'^2)\\
\Psi_2 & =& \Psi_2^0 + (-\eth_0 \Psi_3^0 + 2 \bar\pi_0 \Psi_3^0 + 3 \mu_0 \Psi_2^0) r'+ O(r'^2)\\
\Psi_3 &=& \Psi_3^0 + (-\eth_0 \Psi_4^0 +   \bar\pi_0 \Psi_4^0 + 4
\mu_0 \Psi_3^0) r'+ O(r'^2) \nn
 \ea\ee
%
%
\be\ba{lllll}

\dot \rho_0  &=& \bar\eth_0 \kappa_0
-\bar\kappa_0\bar\pi_0 -\kappa_0\pi_0,  & \dot \sigma_0  =  \eth_0 \kappa_0,\\
%
 \dot\pi_0 &+&  \kappa_0=0,  & \dot\lambda_0 = \bar\eth_0 \pi_0-\pi_0\bar\pi_0
- 2 \lambda_0 \epsilon_0, \\
 \dot \a_0  &-&\bar P  \up{\bar c}\nabla \epsilon_0 =0, & \dot \b_0  - P  \up{c}\nabla \epsilon_0 =0,\\
 \dot \mu_0 &=& \eth_0\pi_0 +\pi_0\bar\pi_0   \\
  &&-2 \mu_0 \epsilon_0 + \Psi_2^0,  \\
\eth_0 \rho_0 &-&\bar\eth_0 \sigma_0 = - \Psi_1^0, &
 \eth_0 \lambda_0 -\bar\eth_0 \mu_0 = - \Psi_3^0, \\
 \Psi_2^0 &=& \bar P  \up{\bar c}\nabla \b_0 - P
\up{c}\nabla \a_0 + \\
&& \a_0\bar\a_0 +
\b_0\bar\b_0 - 2 \a_0\b_0, &
  2 \Im \eth_0
\pi_0= -2 \Im \Psi^0_2, \\
  \kappa_0 &=& -2 P \up{c}\nabla \epsilon_0, & \dot P =0,\\
 && \bar P  \up{\bar c}\nabla \ln P = \b_0 -\bar
 \a_0, \\
%
\dot \Psi_1^0
 &=&  - 3 \kappa_0 \Psi_2^0, & 
 \dot \Psi_2^0
 = 0,\\
 \dot \Psi_3^0
 &-& \bar\eth_0 \Psi_2^0 = 3 \pi_0 \Psi_2^0 -2 \epsilon_0
 \Psi_3^0, \\
 \dot \Psi_4^0
 &-& \bar\eth_0 \Psi_3^0 = -3 \lambda_0 \Psi_2^0 \\&&+ 4 \pi_0 \Psi_3^0- 4 \epsilon_0 \Psi_4^0.
 \nn\ea\ee
where the complex derivative is defined as 
$ \up{c}\nabla:= \frac{\partial}{\partial x^2} +
i\frac{\partial}{\partial x^3}$,
$ P(v,x^k):= \xi^{30}= -i \xi^{40}$ and $ P\up{c}\nabla =\delta_0. $ 

\textbf{Surface gravity: from NEH to WIH}

Here we prove that the surface gravity for a
rotating WIH is also constant. From \textbf{(d)} in p. 46 and complex conjugate of \textbf{(e)}
in p. 46 \footnote{Here the number of NP equations refer to p.
46-p. 50  in \cite{Chandrasekhar}.}, we have
    \bea \dot \pi_0= \frac{d}{d v} (\a_0+\bar\b_0) = 2 \bar P \up{c}\nabla \epsilon_0 = -\bar\kappa_0.\eea
    Using \textbf{(c)} in p. 46, $\dot\pi_0 =-\kappa_0$. It implies $\kappa_0 -\bar\kappa_0
    =0$. Therefore $\kappa_0$ is real. Using \textbf{(b)} in p. 46, we get $\kappa_0
    =0$, i.e., $\delta_0 \epsilon_0=0$ (i.e., $P\up{c}\nabla \epsilon_0 =0$) on NEH.

We make a coordinate choice $r_0=- \frac{1}{\mu_0}$ on the NEH and
it becomes a WIH. This gives $\dot \mu_0 =0$. Applying time
derivative on \textbf{(h)} in p. 46 and using $\kappa_0=0$ from
\textbf{(b)} in p. 49, we then get $\dot\epsilon_0=0$.

It then gives us that $\epsilon_0$ is constant on WIH. So the
surface gravity $\kappa_{(\l)} \hat{=}\Re\epsilon_0$ is
constant on WIH. For NEH, the surface gravity is not necessary
constant.

\textbf{Coordinate transformations near generic isolated
horizons}

We look at the coordinate transformations
on the horizon which  are  similar to those of the Newman-Unti or
BMS group. Under such coordinate transformations, the metric form
is preserved. The metric components can be expanded in terms of $r'$:
\bea g^{vr'}&=& -1, \; \; g^{vv}=g^{vk} =0,  \\
g^{r'r'} &=&  -2 U= -2 \epsilon_0 r' - \epsilon_1 r'^2 + O(r'^3)\\
g^{r'k} &=& - X^k= - X^{0k} - X^{1k} r' + O(r'^2)\\
g^{mn} &=&  -(\xi^m \bar\xi^n + \bar\xi^m\xi^n)= -2 P\bar P
\delta^{mn} + O(r'),
 \eea
where $ k, m, n = 3,4$. We expand the new coordinates $(\tilde{v}, \tilde{r'} ,
\tilde{x^m})$, in terms of $r'$ to obtain
\bea  \tilde{v} &:=& V_0 + V_1 r' + V_2 r'^2 + O(r'^3),\\
\tilde{r'} &:=& R_1 r' + R_2 r'^2+O(r'^3),\\
\tilde{x^m} &:=& K^m_0 + K^m_1 r' + K^m_2 r'^2+O(r'^3).\eea
%
%
We use $g^{\ta\tb} = \frac{\partial x^\ta}{\partial x^c}
\frac{\partial x^\tb}{\partial x^d} g^{cd} $ to transform the metric into the new
coordinates, and then obtain the conditions for the metric
components. From $g^{\one\tk}$, we get the condition
$ \frac{\partial K^k_0}{\partial v} + \frac{\partial
K^k_0}{\partial x^l} X^{0l} + 2 K^k_1 \epsilon_0=0
$
to make $X^{\tn\tk}=0$ in the new coordinate. $R_1$ can be solved from the condition of preserving lowest order of $g^{11}$ and $V_0$ can be integrated from condition of $g^{01}$ (see \cite{Wu2007}). Therefore, when $r$ approaches $r_\Delta$ we have the infinitesimal
coordinate transformation on horizon which is
\be\ba{lll} \tilde{v}&\hat{=}& V_0 = \frac{1}{\epsilon_0} \ln (G(x^k) + e^{\epsilon_0 v})\\
  \tilde{r}' &\hat{=}& R_1 r' = (G(x^k) e^{-\epsilon_0 v} +1)\;  r'\\
  \tilde{x^k} &\hat{=}& K^k_0(v, x^k). \label{coor-trans}
\ea\ee
From the coordinate transformation (\ref{coor-trans}) near
horizon, it gives the asymptotic symmetric group transformation
near a generic isolated horizon. It is similar with the asymptotic symmetric group (the so called BMS
    group or Newman-Unti group) near null infinity.
    When $ \tilde{v}= v + H(x^k), \tilde{x^k} = x^k$, it then defines the analogues of
    \textit{supertranslations}
    which generate different cuts on generic isolated horizons.

\section{Constant spinors for the generic isolated horizons: Frame alignment\label{conspinor}}

In this section, we adopt a similar idea of Bramson's asymptotic frame aligment \cite{Bramson75a} to set up spinor frames on horizon. Firstly, we demand the conditions on spinor frames to be
parallelly transported along the horizon generators $\l^a$
direction on $\Delta$, so
\be\ba{l} \lim_{r'\to 0} D Z_A\,^{\underline{A}}=0, \label{SD}
\ea\ee
and also the conditions of the frames on different generators on
$\Delta$ are:
\be\ba{l} \lim_{r'\to 0} \delta Z_A\,^{\underline{A}}=lim_{r'\to
0}\; \bar\delta Z_A\,^{\underline{A}}=0 . \label{SDelta}\ea\ee
It leads to the six conditions for the constant spinor $\lambda_A$  are
\footnote{We define $Z_A\,^{\underline{A}} = (\lambda_A, \mu_A)$ and
$ \lambda_A = \lambda_1 o_A -\lambda_0 \iota_A$, $\mu_A = \mu_1
o_A - \mu_0 \iota_A$
where
$\lambda_1= \lambda_1^0(v,\theta,\phi) +
\lambda_1^1(v,\theta,\phi) r' + O(r'^2),
\lambda_0=
\lambda_0^0(v,\theta,\phi) + \lambda_0^1(v,\theta,\phi) r' +
O(r'^2) $
and $\lambda_1$ is type $(-1,0)$ and $\lambda_0$ is type $(1,0)$.}
\bea \dot{\lambda_0^0} -\epsilon_0 \lambda^0_0 &=&0,  \textrm{i.e., }\;\; \thorn_0 \lambda_0^0 =0 \label{Framee1}\\
     \dot{\lambda_1^0} +\epsilon_0 \lambda_1^0 &=&\pi_0 \lambda_0^0,  \textrm{i.e., }\;\; \thorn_0 \lambda_1^0 =\pi_0 \lambda_0^0  \label{Framee2}\\
     \eth_0 \lambda_0^0             &=& 0                  \label{Framee3}
\eea
\bea
     \eth_0 \lambda_1^0  - \mu_0 \lambda_0^0 &=& 0       \label{Framee4}\\
     \eth'_0 \lambda_0^0      &=& 0                  \label{Framee5}\\
     \eth'_0 \lambda_1^0  +\sigma'_0 \lambda_0^0 &=& 0.        \label{Framee6}
\eea
To avoid confusion with the spinor
$\lambda_A$, we use another symbol $-\sigma'_0$ to represent $\lambda_0$, i.e., 
the NP shear of $n$.


We use the condition (\ref{Framee1}) and the fact that
$\thorn_0\eth_0 =\eth_0\thorn_0$ on the horizon. Apply $\thorn_0$
on (\ref{Framee3}), we find
\bea 0= \thorn_0 \eth_0 \lambda_0^0 = \eth_0 \thorn_0 \lambda_0^0
=0.\eea
So condition (\ref{Framee1}) and (\ref{Framee3}) are compatible.

Apply $\thorn$ on (\ref{Framee4}) and use condition
(\ref{Framee1}) and (\ref{Framee2}), we have
\bea 0 &=& \Psi^0_2 \lambda_0^0.\eea
Hence the condition (\ref{Framee1}), (\ref{Framee2}) and
(\ref{Framee4}) are not compatible unless $\Psi_2^0=0$.


From the previous analysis, we conclude that \textit{the
compatible frame alignment conditions for the generic isolated
horizon} are (\ref{Framee1}), (\ref{Framee3}) and (\ref{Framee4}).
Equation (\ref{Framee3}) and (\ref{Framee4}) are Dougan-Mason's
holomorphic conditions \cite{Dougan&Mason}. Here we see that the
conditions of the spinor field to be asymptotically constant on NEH
implies the Dougan-Mason holomorphic conditions on the cuts of the
NEH. These equations will be used together with the Nester-Witten
two form to define the quasi-local energy-momentum. The time
related condition (\ref{Framee1}) will tell us how the energy
momentum changes with time along NEH and will be useful to
calculate the energy flux across the horizon.

%
\section{The quasi-local energy-momentum of an isolated horizon}\label{emIH}

By using Nester-Witten two form together with the compatible
constant spinor conditions which are Dougan-Mason's holomorphic
conditions (\ref{Framee3}) and (\ref{Framee4}) for the NEH, 
the quasi-local momentum integral near a NEH is
\be\ba{lll} I(r') &=& -\frac{1}{8\pi} \oint_{S_r}
[\lambda_{0'}\eth \lambda_1 - \lambda_{1}\eth \lambda_{0'} +
\lambda_{0}\eth' \lambda_{1'} - \lambda_{1'}\eth' \lambda_{0}\\
 & &- \lambda_0\lambda_{0'} (\mu+\bar \mu) - \lambda_1\lambda_{1'}
(\rho+\bar\rho)] d S \\
&=&  \frac{1}{4\pi} \oint_S [-\mu_0 \lambda_0^0 \bar\lambda^0_{0'}
+ O(r')]d S. \label{mom-int} \ea\ee
Moreover, the horizon momentum  $P_{\underline{AA'}}$ can
be written as
\bea P_{\underline{AA'}} (S_\Delta) =
I(r_\Delta)\lambda_{\underline{A}}\bar\lambda_{\underline{A'}}
\eea
where $\lambda_{\underline{A}}$ is constant spinor on two surface
of NEH. From the result of the asymptotic expansion for the
generic isolated horizons, we can re-interpret the
\textit{quasi-local energy-momentum integral of the generic
isolated horizons (NEH)} as
\bea I(r_\Delta) = -\frac{1}{4\pi} \oint_S \frac{\Psi^0_2
-\dot\mu_0 + \eth_0 \pi_0 + \pi_0\bar\pi_0}{2 \epsilon_0}
\lambda_0^0 \bar\lambda^0_{0'}
 d S_\Delta \label{quasilocal-RNEH}\eea
where $\Psi_2^0 = M + i L$ and $\eth_0\pi_0 = A -i L$ with $M, L,
A$ are function of $(v,\theta,\phi)$.

\section{News function and conserved quantities of generic isolated horizons}
\label{NEHflux}

In order to match the Kerr solution that its flux
vanishes, we rescale the spinor field. Firstly, the constant
spinors $\lambda^0_0$ and $\lambda^0_1$ are rescaled by using the
following relation
\be\ba{lllll} \tilde{\lambda}^0_0 &=& \lambda^0_0 e^{-\int
\epsilon_0 d v},\;\;
      \tilde{\lambda}^0_{1} &=& \lambda^0_{1} e^{-\int \epsilon_0 d
      v}, 
      \ea\ee
and it yields the new rescaled momentum integral
\bea \tilde{I}(r_\Delta) &=& e^{-2 \int \epsilon_0 d v}
I(r_\Delta)
           = -\frac{1}{4\pi} \oint \mu_0 \tilde{\lambda}^0_0
\tilde{\bar\lambda}^0_{0'} d S_\Delta. \eea
The three compatible conditions (\ref{Framee1}), (\ref{Framee3})
and (\ref{Framee4}) then become
\bea \dot{\tilde{\lambda}}^0_0 =0,\;\; \eth_0
\tilde{\lambda}^0_0=0, \;\; \eth_0 \tilde{\lambda}^0_1 - \mu_0
\tilde{\lambda}^0_0= 0  \eea
where we use $\delta_0 \epsilon_0=0$ from asymptotic expansion and they are still compatible under rescaling.

By using this new rescaling constant spinor frame, we apply the
time derivative on the quasi-local energy-momentum of NEH
(\ref{quasilocal-RNEH}) and thus we get
\bea \dot{\tilde{I}}(r_\Delta) = - \frac{1}{4\pi}\oint \dot\mu_0
\tilde{\lambda}^0_0 \tilde{\bar\lambda}^0_{0'} d S_\Delta \label{NewdotI}\eea
We claim that (\ref{NewdotI}) is \textit{quasi-local energy flux near NEH}. The area will not change for the generic isolated horizons, so the time derivative of the surface area element vanish. Here we can
see the fact that $\dot \mu_0$ is related with the mass loss or
gain, hence it is the \textit{news function of the generic
isolated horizons}.


We now consider the absolute conservation law that $\dot G_m=0$ on
the generic isolated horizons. From \textbf{(b)} in p. 49, we have
$\dot\Psi_2^0=0$, therefore, we can find ten conserve quantities
which are
\be G_m= \int {_2} Y_{2,m} \Psi_2^0 d S .  \;\;\;\;\;
(m=-2,-1....,2)\ee
Here these conserved quantities corresponds to three different
type of generic isolated horizons. Firstly, the most general NEH
does not need to require any stationary. Therefore, we have mass
loss or mass gain from the outgoing radiation along NEH. Secondly,
we have $\dot\mu_0=0$ on WIH. Therefore, there is no gravitational
radiation on WIH From our asymptotic expansion, it further implies
two conditions $\dot\epsilon_0=0, \dot\pi_0=0$. This part is
different from Ashtekar's construction. The most restrict
definition is IH. It further needs the condition
$\dot\lambda_0=-\dot\sigma_0'=0$. This is a fully stationary case
in our construction.

\section{Conclusions}

We work out the coordinate transformations which carry out the
asymptotic symmetric group near the generic isolated horizon.
Asymptotically constant spinors can be used to define the
quasi-local energy-momentum of the horizon. Searching for the
compatible conditions of constant spinors of horizons offers us a
way to chose for the reference frame when measuring these
quasi-local quantities. We find that the news function exists only
for NEH. It indicates the radiation outside the equilibrium
black hole. The radiation will not cross the horizon, therefore the
area will not increase. The news function of NEH will vanish while
we make a special choice of affine parameter $r_\Delta=
-\frac{1}{\mu_0}$. This result refers to that a generic NEH admits
an unique $[\l]$ such that $(\Delta,[\l])$ is a WIH on which the
incoming expansion $\mu_0$ is time independent \cite{Ashtekar99b}.
The conserved quantities of the generic isolated horizon is easily
shown from the equations of the asymptotic expansion. For a
stationary case, it corresponds to mass and angular momentum.

\acknowledgments
 YHW would like to thank her host Prof Chung-Ming Ko and this work was
supported by the National Science Council (NSC) of
Taiwan, Grant No. NSC 96-2811-M-008-056 (YHW) and No. NSC 096-2811-M-008-040 (CHW).


\end{document}